\documentstyle[12pt]{article}
\begin{document}
\vskip 2.00cm
\title{\bf The constant background bag model of the hadron}
\author{Sh. Mamedov\thanks{Email: sh$_-$mamedov@yahoo.com \hspace{2mm} \&
\hspace{2mm}  shahin@theory.ipm.ac.ir }\\
{\small {\em Institute for Studies in Theoretical Physics and
Mathematics (IPM),}} \\
{\small {\em P.O.Box 19395-5531, Tehran, Iran}}\\
{\small {\em and }} \\
{\small {\em High Energy Physics Lab., Baku State University, }}\\
{\small {\em Z.Khalilov str.23, Baku 370148, Azerbaijan}}}
\maketitle
\begin{abstract}
\noindent We have constructed the bag model having a central constant color field. The motion of the quark is studied in this bag and the Dirac equation is solved for it. The energy spectrum found has a branching due to the interaction of the quarks with the color background. It is pointed out that this model can be applied for taking into account, in mass spectrum of the hadrons, the coupling of the constituent quarks with the gluon condensation as the interaction with the color background.
\end{abstract}

\hspace{15mm} \newline {\Large {\bf Introduction}} \vspace{4mm} \newline 
\noindent There are a constituent quark model and some bag models describing
hadron states by means of the dynamics and kinematics of the constituent quarks
[1-4]. These models take into account the gluon and quark condensation
existing in the QCD vacuum [5] as a medium in which the constituent quarks move and
the interaction of these quarks with condensates are reduced acquiring
additional mass by them. The vacuum average values of the chromofield components 
$\langle {\cal E}^a\,{\cal E}^a\rangle $ and $\langle {\cal H}^a\,{\cal H}^a\rangle $ of the gluon condensate in the
QCD vacuum were estimated in [5-7]. These estimations are used for the study
of vacuum energy and structure of the bag models, which connected with the bag
constant [8,9]. Here we aim to consider the gluon condensate as a color
background field and take into account the coupling of the constituent quarks
with it as an interaction with the external color filed. Naturally, such
a coupling will rise in the constituent quarks energy spectrum which is not consisting in
only acquiring additional mass term and will result in a splitting of this spectrum. In order to provide the rotational invarince we propose that spatial components of this background are equal and constant for simplicity. The background field thus defined will play the role of
a central constant color force. In order to realize our proposals we shall construct
a model like Bogolubov's bag model [3], but having a constant color field in
the center. According to this model the colored constituent spinor particle
moves in the constant central color field and its motion is limited by
an infinite spherical well. The condition of an infinite spherical well is imposed in order to
ensure the confinement property of the colored constituents. The Dirac equation
should be solved for this model under this boundary condition on wave
functions of the constituent quark. For this aim we can use the constant
non-abelian vector potentials found in [10] and the approach used for solving
the Dirac equation in the field given by such a kind of potentials [11]. This will
give us the distribution of constituent quarks inside the bag and their energy
spectrum, which will contain the contribution of its color and spin interactions
with the background field. Finally, the components of external field can be
identified with the ones of the gluon condensate and the estimations for the color
components of the gluon condensation $\langle {\cal E}^a\,{\cal E}^a\rangle $ and $\langle
{\cal H}^a\,{\cal H}^a\rangle $ can be setted in the energy spectrum found for the quark. This
will make this spectrum useful for a comparison with the spectra of hadron states.

\section{The Dirac equation}

For a study of the problems in a constant non-abelian background field it is convenient to introduce the constant non-commuting vector potentials%
\footnote{See [11] and references therein.}. Let us choose the vector potentials 
$A_\mu ^{(a)}$ in the framework of the $SU_c(3)$ color symmetry group in the
following way: 
\begin{equation}
\label{1}
\begin{array}{c}
\ A_0^{(a)}=
\sqrt{\tau _1},\ A_j^{(a)}=\sqrt{\tau }\delta _{ja}\ \left( j=1,2,3\right) \
for\ a=1,2,3 \\ A_\mu ^{(a)}=0\ for\ a=4,5,6,7,8,
\end{array}
\end{equation}
where $\tau ,\tau _1$ are constants and $\delta _{ia}$ is the Kronecker symbol.
The corresponding field strength tensor $F_{\mu \nu }^{(a)}=gf^{abc}A_\mu ^{(b)}A_\nu
^{(c)}$ has the following chromoelectric $({\cal E}_j^{(a)})$ and chromomagnetic $%
({\cal H}_j^{(a)})$ components: 
\begin{equation}
\label{2}
\begin{array}{c}
{\cal E}_x^{(1)}=0,\ {\cal E}_x^{(2)}=g
\sqrt{\tau \tau _1},\ {\cal E}_x^{(3)}=-g\sqrt{\tau \tau _1};\ {\cal H}_x^{(1)}=g\tau ,\
{\cal H}_x^{(2)}=0,\ {\cal H}_x^{(3)}=0,\  \\ {\cal E}_y^{(1)}=-g
\sqrt{\tau \tau _1},\ {\cal E}_y^{(2)}=0,\ {\cal E}_y^{(3)}=g\sqrt{\tau \tau _1};\
{\cal H}_y^{(1)}=0,\ {\cal H}_y^{(2)}=g\tau ,\ {\cal H}_y^{(3)}=0,\ \  \\ \ {\cal E}_z^{(1)}=g\sqrt{\tau
\tau _1},\ {\cal E}_z^{(2)}=-g\sqrt{\tau \tau _1},\ {\cal E}_z^{(3)}=0;\ {\cal H}_z^{(1)}=0,\
{\cal H}_z^{(2)}=0,\ {\cal H}_z^{(3)}=g\tau .\ 
\end{array}
\end{equation}
Here $g$ is the color interaction constant. All other color components of $%
F_{\mu \nu }^{(a)}$ are zero for $a=4,5,6,7,8.$ From (2) we see that the field
(1) has equal magnitude of spatial components ${\cal E}_j^2={\cal E}_j^{(a)}{\cal E}_j^{(a)}=2g^2%
\tau \tau _1\ $and ${\cal H}_j^2={\cal H}_j^{(a)}{\cal H}_j^{(a)}=g^2\tau ^2,$ which are constant
as well. So, in ordinary space the strength vectors of the chromomagnetic and
chromoelectric fields are%
\footnote{We assume the chromoelectric field has negative projections in color space.}%
: 
\begin{equation}
\label{3}
{\overrightarrow {\cal H}}
=\sqrt{3} g \tau \,{\overrightarrow n},\ 
{\overrightarrow {\cal E}}=g \sqrt{6\tau \tau _1}\,{\overrightarrow n},
\end{equation}
where ${\overrightarrow n}$ is the unit radius vector in ordinary space.

The Dirac equation for a colored particle minimally coupled with the
external color field (1) can be written as follows:

\begin{equation}
\label{4}\left( \gamma ^\mu P_\mu -M\right) \Psi =0, 
\end{equation}
where $P_\mu =p_\mu +gA_\mu =p_\mu +gA_\mu ^{(a)}\frac{\lambda ^a}2$ , and the $%
\lambda ^a$ are Gell-Mann matrices describing the particle's color spin. Equation
(4) can be written in terms of the Majorana spinors $\phi $ and $\chi $ , 
$$
\Psi =
\left( 
\begin{array}{c}
\phi \\ 
\chi 
\end{array}
\right) ,
$$ 
in the more suitable form for us: 
\begin{equation}
\label{5}\left( \sigma ^jP_j\right) ^2\psi =\left( P_0^2-M^2\right) \psi , 
\end{equation}
where the Pauli matrices $\sigma _i$ describe a particle's spin. Here and
afterwards $\psi $ means $\phi $ or $\chi .$ The two spin components of the Majorana
spinors $\psi =\left( 
\begin{array}{c}
\psi _1 \\ 
\psi _2 
\end{array}
\right) $ transform under the fundamental representation of color group $%
SU_c(3).$ That means that each spin component of the wave function $\psi _{1,2}$ has
three color components describing the color states of a particle:

$$
\psi _{1,2}=\left( 
\begin{array}{c}
\psi _{1,2}^{(1)} \\ 
\psi _{1,2}^{(2)} \\ 
\psi _{1,2}^{(3)} 
\end{array}
\right) . 
$$

Writing down the expressions of $P_\mu $ and $A_\mu ^{(a)}$ in equation (5)
we get an explicit form: 
\begin{equation}
\label{6}\left( {\overrightarrow p}^2+M^2+\frac{3g^2\tau }4+g\tau ^{\frac
12}\lambda ^ap^a-\frac{g^2\tau }2\sigma ^a\lambda ^a\right) \psi =\left(
E^2-g\tau _1^{\frac 12}\lambda ^aE+\frac{3g^2\tau _1}4\right) \psi ,
\end{equation}
where $E$ is the energy of particle%
\footnote{Since the field (1) does not depend on time, the states are stationary.}%
. Equation (6) turns into following system of differential equations for the color components $\psi _{1,2}^{\left( a\right) }$:
\begin{eqnarray}
\label{7}
\left\{
\begin{array}{rll}
\left( A+g\tau ^{\frac 12}{\cal P}_3-\frac 12g^2\tau \right) \psi_1^{(1)}+g\tau
^{\frac 12}\left( {\cal P}_1-i{\cal P}_2\right) \psi _1^{(2)}&=&0  \\ 
\left( A-g\tau ^{\frac 12}{\cal P}_3+\frac 12g^2\tau \right) \psi_1^{(2)}+g\tau
^{\frac 12}\left( {\cal P}_1+i{\cal P}_2\right) \psi _1^{(1)}&=&g^2\tau \psi _2^{(1)} 
\\ 
\left( 
{\overrightarrow p}^2+M^2\right) \psi _1^{(3)}&=&E^2\psi _1^{(3)} \\ \left(
A+g\tau ^{\frac 12}{\cal P}_3+\frac 12g^2\tau \right) \psi _2^{(1)}+g\tau
^{\frac 12}\left( {\cal P}_1-i{\cal P}_2\right) \psi _2^{(2)}&=&g^2\tau \psi_1^{(2)} 
\\ 
\left( A-g\tau ^{\frac 12}{\cal P}_3-\frac 12g^2\tau \right) \psi_2^{(2)}+g\tau
^{\frac 12}\left( {\cal P}_1+i{\cal P}_2\right) \psi _2^{(1)}&=&0   \\ 
\left( {\overrightarrow p}^2+M^2\right) \psi _2^{(3)}&=&E^2\psi_2^{(3)}\,, 
\end{array}
\right.
\end{eqnarray}
where the operators $A$ and ${\cal P}_j$ denote $A={\overrightarrow p}^2+M^2+\frac 34g^2
\left( \tau -\tau _1\right) -E^2,\ {\cal P}_j=p_j+\sqrt{\frac{%
\tau _1}\tau }E.$ The equations in the system (7) mix the different states $\psi _{1,2}^{\left( a\right) }$ and we need separeted equations for each of these states. From the system (7) we find that equations for all states $\psi _{1,2}^{(i)}$ $\left( i=1,2\right) $ have the same form:%
\begin{equation}
\label{8}\left[ \left( A-\frac{g^2\tau }2\right) ^2-g^2\tau {\overrightarrow {\cal P}}^2\right] 
\left[ \left( A+\frac{g^2\tau }2\right) ^2-g^2\tau \left( 
{\overrightarrow {\cal P}}^2+g^2\tau \right) \right] \psi _{1,2}^{(i)}=0,
\end{equation}
which possesses rotational invariance. Since the operators $A$ and ${\cal P}_i$
commute, the operator in first square bracket commutes with the second one.
This allows us to divide the (8) into two 
equations, both keeping rotational invariance: 
\begin{equation}
\label{9}\left[ \left( A-\frac{g^2\tau }2\right) ^2-g^2\tau {\overrightarrow {\cal P}}^2\right] \psi _{1,2}^{(i)}=0,
\end{equation}
\begin{equation} 
\label{10}\left[ \left( A+\frac{g^2\tau }2\right) ^2-g^2\tau \left( 
{\overrightarrow {\cal P}}^2+g^2\tau \right) \right] \psi _{1,2}^{(i)}=0.
\end{equation}
Equations (9) and (10) can be solved separetaly and the set of solutions (8) will consist of the solutions of (9) and (10). Let
us consider equation (9). Acting on this equation by the operator $
{\overrightarrow {\cal P}}^2$ we get the same equation for the function $\xi
_{1,2}^{(i)}={\overrightarrow {\cal P}}^2\psi _{1,2}^{(i)}$ as for $\psi
_{1,2}^{(i)}$, i.e.:%
$$
\left[ \left( A-\frac{g^2\tau }2\right) ^2-g^2\tau {\overrightarrow {\cal P}}^2\right] 
\xi _{1,2}^{(i)}=0. 
$$
This means the functions $\xi _{1,2}^{(i)}$ and $\psi _{1,2}^{(i)}$ differ
only by a constant multiplier $k^{\prime 2}:\xi _{1,2}^{(i)}=k^{\prime
2}\psi _{1,2}^{(i)}$ or 
\begin{equation}
\label{11}{\overrightarrow {\cal P}}^2\psi _{1,2}^{(i)}=k^{\prime 2}\psi
_{1,2}^{(i)}.
\end{equation}
In other words, since the operator ${\overrightarrow {\cal P}}^2$ commutes with
the square bracket operator in (9), they have the same set of eigenfunctions 
$\psi _{1,2}^{(i)}.$ The same claim is order for the operator $
{\overrightarrow p}^2:$%
\begin{equation}
\label{12}{\overrightarrow p}^2\psi _{1,2}^{(i)}=-\nabla ^2\psi
_{1,2}^{(i)}=k^2\psi _{1,2}^{(i)}.
\end{equation}
Thus, we can solve (12) instead of (9). Obviously,
(12) keeps the rotational invariance property of equivalent
equation (9) and so is easily solved in a spherical coordinate system using the separation ansatz [12]: 
\begin{equation}
\label{13}\psi _{1,2}^{(i)}\left( {\overrightarrow r}\right) =R(r)\cdot
Y_l^m\left( \theta ,\varphi \right) .
\end{equation}
Here $r=\sqrt{x^2+y^2+z^2},$ $l$ and $m$ are the orbital angular momentum
and chromomagnetic quantum numbers, $\theta ,\varphi $ are the polar and
azimuthal angles, respectively. The spherical functions $Y_l^m\left( \theta
,\varphi \right) $ are expressed by means of the Legendre polynomials $%
P_l^{\mid m\mid }\left( \cos \theta \right) :$%
\begin{equation}
\label{14}Y_l^m\left( \theta ,\varphi \right) =\sqrt{\frac{\left(
2l+1\right) }{4\pi }\frac{\left( l-\mid m\mid \right) !}{\left( l+\mid m\mid
\right) !}}P_l^{\mid m\mid }\left( \cos \theta \right) e^{im\varphi },
\end{equation}
and define the $``s"$, $``p"$, $``d"$, $``f"$, $\cdots $ orbitals well-known in quantum mechanics. The equation for the radial part $R(r)$ is as used in many quantum
mechanical problems possessing rotational invariance [12,11]: 
\begin{equation}
\label{15}\frac{d^2}{dr^2}R(r)+\frac 2r\frac d{dr}R(r)+\left( k^2-\frac{%
l\left( l+1\right) }{r^2}\right) R(r)=0.
\end{equation}
With the notations $Q\left( r\right) =\sqrt{r}R(r)$ (15) turns into Bessel's
equation for $Q\left( r\right) :$ 
\begin{equation}
\label{16}Q^{\prime \prime }\left( r\right) +\frac 1rQ^{\prime }\left(
r\right) +\left( k^2-\frac{\left( l+\frac 12\right) ^2}{r^2}\right) Q\left(
r\right) =0.
\end{equation}
The function $R(r)$ must be finite on $r\rightarrow 0.$ This means that for the
solution of (15) we should choose the Bessel function of the first kind: 
\begin{equation}
\label{17}R_l(r)=\frac{C_l}{k\sqrt{r}}J_{l+1/2}\left( kr\right) .
\end{equation}
Thus, we conclude that states obeying (9) are the $``s"$, $``p"$, $``d"$, $``f"$, $\cdots $ orbitals corresponding to the
different values of the quantum numbers $l$ and $m$ and the motion of the constituent quarks in any color and spin state takes place
on these orbitals, which are the same for all these states. In other words, the angle
distribution of quarks inside bag is the same for any color and spin state and the same, for instance, as ones of electrons
in atoms%
\footnote{Since constituents in this model are considered to be non-interacting with each other, for the two-particle case we have the angle distribution of quarks in mesons. This distribution will have the same shape as ones in two-electron atom. In this sense mesons like hydrogen atom, differing only by the radial distribution
in the constant field approximation.}. The radial distribution
of these quarks is the same as ones of freely moving particles enclosed in a
sphere [12]. Since, the external field (1) does not depend on $r,$ we have
obtained the same expression for the solutions $\psi _{1,2}^{(i)}\left( 
{\overrightarrow r}\right) $ as for a freely moving particle enclosed in a
sphere, differing only by the expression of the $k^2$ constant.

\section{The energy spectrum}

Using (3) the operator ${\overrightarrow {\cal P}}$ can be written in the
following form:%
$$
{\overrightarrow {\cal P}}={\overrightarrow p}+\frac E{\sqrt{6}g\tau }
{\overrightarrow {\cal E}}. 
$$
Then the action of ${\overrightarrow {\cal P}}^2$ operator will be 
\begin{equation}
\label{18}{\overrightarrow {\cal P}}^2\psi _{1,2}^{(i)}={\overrightarrow p}^2\psi
_{1,2}^{(i)}+\frac{2E}{\sqrt{6}g\tau }\left| {\overrightarrow p}\psi
_{1,2}^{(i)}\right| \left| {\overrightarrow {\cal E}}\right| \cos \alpha +\frac{\tau
_1}\tau E^2\psi _{1,2}^{(i)},
\end{equation}
where $\left| {\overrightarrow p}\psi _{1,2}^{(i)}\right| $ means the directional derivative [13]:%
$$
\left| {\overrightarrow \nabla}\Phi \right| =\sqrt{\left( \frac{\partial
\Phi }{\partial x}\right) ^2+\left( \frac{\partial \Phi }{\partial y}\right)
^2+\left( \frac{\partial \Phi }{\partial z}\right) ^2}. 
$$
Here $\alpha $ is the angle between the momentum vector ${\overrightarrow \nabla}\psi $ and the chromoelectric field vector ${\overrightarrow {\cal E}}.$
Since the particle periodically moves on orbitals the angle $\alpha $ varies
in symmetric limits. That means the average value of $\cos \alpha $ during
one period is zero. Actually, the term, which contain $\cos \alpha $ in
(18) is proportional to the work done by chromoelectric field on the particle
in its motion in this field. It is easily seen the net work of this
field during one period is zero, while its momentary value is not zero%
\footnote{We observe the same situation as in an electron's motion in the field of nucleus.}%
. So, if we average (9) and (10) over the time during one
period, the term proportional to $\cos \alpha $ will drop out. We shall find
the average value of energy spectrum%
\footnote{For "s" orbitals $\cos \alpha $=0, since the momentum and chromofield vectors are perpendicular on every moment in time. So, the momentary value of energy coincide with its average value for these orbitals.}%
. Therefore, the average value of (18) during one period is equal to: 
\begin{equation}
\label{19}{\overrightarrow {\cal P}}^2\psi _{1,2}^{(i)}=-\nabla ^2\psi
_{1,2}^{(i)}+\frac{\tau _1}\tau E^2\psi _{1,2}^{(i)}=k^2\psi _{1,2}^{(i)}+
\frac{\tau _1}\tau E^2\psi _{1,2}^{(i)}=k^{\prime 2}\psi _{1,2}^{(i)}.
\end{equation}
If we take (19) and (12) into account, we obtain from (9) the following equation for
the constant $k^2$:%
$$
\left( k^2\right) ^2+2k^2\left( M^2-E^2+\frac 14g^2\left( \tau -3\tau
_1\right) -\frac 12g^2\tau \right) +\left( M^2-E^2+\frac 14g^2\left( \tau
-3\tau _1\right) \right) ^2-g^2\tau _1E^2=0, 
$$
from which one finds the relation between the constant $k^2$ and the energy
spectrum $E^2:$%
\begin{equation}
\label{20}\left( k^2\right) _{1,2}=\left( \sqrt{E^2\left( 1+\frac{\tau _1}%
\tau \right) -M^2+\frac 34g^2\tau _1}\pm \frac 12g\tau ^{\frac 12}\right)
^2-E^2\frac{\tau _1}\tau .
\end{equation}

Now we should impose the boundary condition meaning the confinement
property of constituent quark on its wave function. Since (12) is the Laplace equation we can impose the Dirichlet boundary
condition $\psi _{1,2}^{(i)}\left( r=r_0\right) =0$ or $R_l(r_0)=0,$ which
means that we enclose the particle motion by a sphere with a radius $r_0.$
Here $r_0$ agrees with the half of hadron size. This boundary condition
establishes the following relation between the values of $k$ and the zeros $%
\alpha _l^{(N)}$ of the Bessel function $J_{l+1/2}\left( kr\right) :$%
\begin{equation}
\label{21}kr_0=\alpha _l^{(N)}
\end{equation}
which means the quantization of the $k$ values. According to the relation (20)
the energy spectrum of particle is quantized as well. Plugging (21) in (20), we
find the first two branches of the quantized energy levels of the spectrum: 
\begin{equation}
\label{22}\left( E_l^{(N)}\right) _{1,2}^2=\left( \sqrt{\left( 1+\frac \tau
{\tau _1}\right) \left( \frac{\alpha _l^{(N)}}{r_0}\right) ^2+M^2+\frac
14g^2\left( \tau -2\tau _1\right) }\mp \frac 12g\tau _1^{\frac 12}\right)
^2-\frac \tau {\tau _1}\left( \frac{\alpha _l^{(N)}}{r_0}\right) ^2.
\end{equation}
In (22) and (21) $N$ labels the sequence of zeros of the Bessel function$\
N=1,2,3.....$ and is called the radial quantum number [12]. So, the value of
angular momentum $l$ determines the series of the energy spectrum of the particle
and the radial quantum number $N$ determines the energy levels in this series.
Thus, the finiteness condition imposed on the motion of the particle, because of its
confinement property, of course, leads to the quantization of the energy
spectrum.

As is seen from (21) the constant $k$ gets the same values for the various
branches $\left( k^2\right) _{1,2}:$ 
$$
k_l^{(N)}=\frac{\alpha _l^{(N)}}{r_0}. 
$$
The radius $a$ of the turning point of the particle could be found from (15)
using the maximum condition on the radial function $R^{\prime }(a)=0$ and is equal
to:%
$$
a_l^{(N)}=r_0\frac{\sqrt{l\left( l+1\right) }}{\alpha _l^{(N)}}. 
$$
These radia, as the energy spectrum and $k$, are quantized and are determined
by the quantum numbers $l$ and $N$ and do not depend on the field intensities.
So, there is no difference in values of the radia with the case of motion in a pure
chromomagnetic field [11].

We have solved (9) and found the corresponding two branches of
the energy spectrum. In the same manner we can solve  (10) too. For this equation the function $\xi _{1,2}^{(i)}$ denotes 
$\xi _{1,2}^{(i)}=\left({\overrightarrow {\cal P}}^2+g^2\tau \right) \psi
_{1,2}^{(i)}$ and the following equations are equivalent to (10): 
\begin{equation}
\label{23}\left( {\overrightarrow p}^2+g^2\tau \right) \psi
_{1,2}^{(i)}=K^2\psi _{1,2}^{(i)},\ \left({\overrightarrow {\cal P}}^2+g^2\tau
\right) \psi _{1,2}^{(i)}=K^{\prime 2}\psi _{1,2}^{(i)}.
\end{equation}
The relation between the expressions of $k$ and $K$ is obvious $%
K^2=k^2+g^2\tau ,K^{\prime 2}=k^{\prime 2}+g^2\tau $ and the relation
between $K$ and $K^{\prime }$ is analogous to (19): 
\begin{equation}
\label{24}K^{\prime 2}=K^2+\frac{\tau _1}\tau E^2.
\end{equation}
Taking (23) and (24) into account, we get from (10) an algebraic relation
between $k^2$ and $E^2$: 
\begin{equation}
\label{25}\left( k^2+M^2+\frac 34g^2\left( \tau -\tau _1\right) -E^2+\frac{%
g^2\tau }2\right) ^2-g^2\tau \left( k^2+\frac{\tau _1}\tau E^2+g^2\tau
\right) =0.
\end{equation}
Of course, the solution of (23) is (13) with (14) and (17). The  Dirichlet boundary
condition $R_l(r_0)=0$ applies to (10) too. In the result of
the quantization (21), from (25) we find the other two branches of the energy
spectrum: 
\begin{equation}
\label{26}
\begin{array}{c}
\left( E_l^{(N)}\right) _{3,4}^2=\left( 
\sqrt{\left( 1+\frac \tau {\tau _1}\right) \left( \left( \frac{\alpha
_l^{(N)}}{r_0}\right) ^2+g^2\tau \right) +M^2+\frac 14g^2\left( \tau -2\tau
_1\right) }\mp \frac 12g\tau _1^{\frac 12}\right) ^2 \\ -\frac \tau {\tau
_1}\left( \left( \frac{\alpha _l^{(N)}}{r_0}\right) ^2+g^2\tau \right) .
\end{array}
\end{equation}
As seen from (22) and (26), the energy spectrum of the quark contains the
contribution of quark's interaction with the background field and
this energy spectrum is quantized due to the finiteness of motion of quark.
Another difference with the existing bag models [1-4] is the branching of
the energy spectrum, which is the result of the color and spin interaction of the particle with the background field and is not determined by these quantum
numbers. We have four color and spin states $\psi_{1,2}^{(i)}$ of the quark and four energy branches $\ (E_l^{(N)})_{1,2,3,4}$, but we do not have  one-to-one correspondence of these states and the branches. All these states 
could get energy from the any of these branches. This is the difference between the branching of spectrum and the splitting levels. In fact, here we have obtained the branching of
energy spectrum instead of splitting levels. Such a branching takes place in
a pure chromomagnetic field case [11,14,15] and is connected with the existence
of conserved operator, which contains the color spin $\frac{\lambda ^a}2$ and
the spin $\frac{\sigma _j}2$ operators. There is no need to introduce the color
states in bag models, which do not
deal with the color interaction of particle with the background,  and so, this branching of energy spectrum does not occur in those models. Note that the energy
levels in these branches are determined only by the quantum numbers $l$ and $N$, and so
they are $2l+1$ fold degenerate in the quantum number $m,$ which occurs on
motion in any central field.

The states $\psi _{1,2}^{(3)}$ correspond to the states of a colorless
particle, and (12) for them $\left( k^2=E^2-M^2\right) $ has the
solution (13) with (15) and (17) too. The energy spectrum of these states
is: 
\begin{equation}
\label{27}\left( E_l^{(N)}\right) _{5,6}^2=\left( \frac{\alpha _l^{(N)}}{r_0}%
\right) ^2+M^2. 
\end{equation}

As an application of this model we can propose the central color field is the background field of gluon condensation in QCD vacuum. Then we can use the estimation for vacuum average values of chromoelectric and chromomagnetic
field intensities of this condensation [6,7]:%
$$
\langle 0\mid g^2{\overrightarrow {\cal E}}^a{\overrightarrow {\cal E}}^a\mid 0\rangle
\simeq -\left( 700\ Mev\right) ^4,\ \langle 0\mid g^2{\overrightarrow {\cal H}}^a
{\overrightarrow {\cal H}}^a\mid 0\rangle \simeq \left( 700\ Mev\right) ^4. 
$$
Identifying these estimates with the field intensities in our model $
g^2{\overrightarrow {\cal E}}^2=6g^4\tau \tau _1\ $and $g^2{\overrightarrow {\cal H}}^2=3g^4\tau ^2
$, we see that constant $\tau _1$ should be taken equal to $\tau
_1=-\frac 12\tau $. This enable us to evaluate the $g^2\tau$ constant 
$g^2\tau \cong \frac 1{\sqrt{3}}\left( 700\
Mev\right) ^2$. Then the energy spectra (22) and (26) simplify as follows: 
\begin{equation}
\label{28}
\begin{array}{c}
\left( E_l^{(N)}\right) _{1,2}^2=2\left( 
\frac{\alpha _l^{(N)}}{r_0}\right) ^2-\left( \sqrt{\left( \frac{\alpha
_l^{(N)}}{r_0}\right) ^2-\frac 12g^2\tau -M^2}\mp \frac 1{2\sqrt{2}}g\tau
^{\frac 12}\right) ^2, \\ \left( E_l^{(N)}\right) _{3,4}^2=2\left( \frac{%
\alpha _l^{(N)}}{r_0}\right) ^2+2g^2\tau -\left( \sqrt{\left( \left( \frac{%
\alpha _l^{(N)}}{r_0}\right) ^2+g^2\tau \right) -\frac 12g^2\tau -M^2}\mp
\frac 1{2\sqrt{2}}g\tau ^{\frac 12}\right) ^2.
\end{array}
\end{equation}
In order to estimate quark's energy using the spectra (28) and (27) we can set
in them $r_0=R\simeq 1,7\ fm$ [3] or another estimation for hadron's size.
Since constituent quarks in the framework of this model are considered
non-interacting with each other, the energy of any two- or three-quark system will be the
sum of the energy of separate quarks. As an examination of this model, the total energy of the bag can be compared with the corresponding hadron state. It should be noted that the selection rule $\Delta m=0,\pm 1$ and $\Delta l=\pm 1$ for quantum transitions between
the energy levels, which exists in central field problems [12], occurs
for this problem as well and is useful for a calculation of the energy emitted by
the excited bag. There is not preferred rule on the energy branches for this calculation. A comparison of the energies emitted by bag and by the hadron could serve as a another verification of the role of the gluon condensate as constant central color field.

\begin{center}
{\large {\bf Acknowledgements}}
\end{center}


The author thanks IPM for financing his visit to this institute, where part
of the present work has been done. He acknowledges Prof. F. Ardalan and
members of string theory group for useful discussions in seminars and
for hospitality. The author also thanks the referee of [11], whose report had
stimulated the present research and Prof. S. Parvizi for reading the manuscript and for useful dicussion.

\end{document}